\documentclass[12pt]{article}

\usepackage{amsmath,amssymb,amsfonts}
\usepackage[paper=letterpaper,margin=0.9in]{geometry}

\usepackage{graphicx}

\renewcommand{\thefootnote}{\fnsymbol{footnote}}

\newcommand{\be}{\begin{equation}}
\newcommand{\ee}{\end{equation}}
\newcommand{\bea}{\begin{eqnarray}}
\newcommand{\eea}{\end{eqnarray}}
\newcommand{\ba}{\begin{array}}
\newcommand{\ea}{\end{array}}

\newcommand{\comment}[1]{}

\def\ffract#1#2{\raise .3 em\hbox{$\scriptstyle#1$}\kern-.25em/
                \kern-.2em\lower .2 em \hbox{$\scriptstyle#2$}}

\def\part#1#2{{\partial#1\over\partial#2}}

\begin{document} 

\rightline{\tt astro-ph/0801.4023}

\rightline{VPI-IPNAS-08-04}

\vskip0.5truein

\centerline{\bf \Large
Transient Pulses from Exploding Primordial Black Holes}
\centerline{\bf \Large as a Signature of an Extra Dimension}

\vskip0.5truein

\centerline{\bf Michael Kavic,$^{1,2}$\footnote{kavic@vt.edu}{\  }
John H. Simonetti,$^{1,2}$\footnote{jhs@vt.edu}{\ }
Sean E. Cutchin,$^{1,2}$\footnote{scutchin@vt.edu}}

\centerline{\bf Steven W. Ellingson,$^{1,3}$\footnote{ellingson@vt.edu}{\ } and
Cameron D. Patterson$^{1,3}$\footnote{cdp@vt.edu}}

\setcounter{footnote}{0}
\renewcommand{\thefootnote}{\arabic{footnote}}


%
%
%
\vskip0.5truein
\centerline{
{\it $^{1}$Institute for Particle, Nuclear and Astronomical Sciences, Virginia Tech.}}
{\it \centerline{$^{2}$Department of Physics, Virginia Tech, Blacksburg, Virginia 24061, USA.}}
{\it\centerline{$^{3}$Bradley Department of Electrical and Computer Engineering,}} {\it \centerline{Virginia Tech, Blacksburg, Virginia 24061, USA.}}

\vskip0.5truein

\begin{abstract}

An evaporating black hole in the presence of an extra
spatial dimension would undergo an explosive phase of evaporation. We
show that such an event, involving a primordial black hole, can produce
a detectable, distinguishable electromagnetic pulse, signaling the existence of an extra
dimension of size $L\sim10^{-18}-10^{-20}$~m. We derive a generic
relationship between the Lorentz factor of a pulse-producing ``fireball"
and the TeV energy scale. For an ordinary toroidally compactified extra
dimension, transient radio-pulse searches probe the electroweak energy
scale ($\sim$0.1~TeV), enabling comparison with the Large Hadron
Collider. 

\end{abstract}

\newpage
\section{Introduction}

A new generation of radio telescopes will search for transient pulses
from the universe \cite{LWA,MWA,LOFAR,ETA1,ETA2}. Such searches,
using pre-existing data, have recently found surprising pulses of
galactic and extragalactic origin \cite{McLaughlin,Lorimer,Vachaspati}.
While the results will be of obvious astrophysical importance, they
could also answer basic questions in physics which are difficult to
address. In particular, as we will discuss here, searches for transient
pulses from exploding primordial black holes (PBHs) can yield evidence
of the existence of an extra spatial dimension, and explore
electroweak-scale physics.\footnote{The existence of primordial black
holes is an open question. However, there exist models of the early
universe which produce large numbers of primordial black holes and are
consistent with all current observational data (see, for example
\cite{kaw}).} The potential impact could be timely and cut across many
areas of investigation. For example, the Large Hadron Collider (LHC) is
poised to investigate electroweak-scale physics, and may also yield
evidence of the existence of extra spatial dimensions.
Also, intensive work on the unification of quantum mechanics and
gravitation has yielded insightful theoretical advances,
often requiring extra spatial dimensions \cite{GSW}, yet there is little
experimental observation which gives feedback on this proposed
phenomenon. Furthermore, mapping of the anisotropies in the cosmic
microwave background radiation has enabled ``precision cosmology,'' yet
searches for PBHs, which would explore smaller scale primordial
irregularities (a source of PBHs), would be valuable
\cite{Carr}. Searches for transient pulses from exploding
primordial black holes can provide information impacting all of these
areas of investigation, which at first glance appear unrelated, but are
intimately connected.

The defining relation governing the Hawking evaporation of a black hole
\cite{Hawking} is 
\be
T = \frac{\hbar c^3}{8\pi G k} \frac{1}{M},
\label{tem}
\ee
for mass $M$ and temperature $T$. The power emitted by the black hole is
\be
P \propto \frac{\alpha(T)}{M^2},
\label{pow}
\ee
where $\alpha(T)$ is the number of particle modes available.
Equations~(\ref{tem}) and (\ref{pow}), along with an increase in the
number of particle modes available at high temperature, leads to the
possibility of an explosive outburst as the black hole evaporates its
remaining mass in an emission of radiation and particles.\footnote{The
behavior of the evaporation process, as the Planck mass is reached, is
not certain \cite{Bee3}. However, the details of this late stage of evaporation 
will not alter the analysis presented here.} PBHs of sufficiently
low mass would be reaching this late stage now \cite{Carr}.
Searches for these explosive outbursts have traditionally focused on
$\gamma$-ray detection \cite{Halzen}. However, Rees noted that exploding
primordial black holes could provide an observable coherent radio pulse
that would be easier to detect \cite{Rees}.

Rees \cite{Rees} and Blandford \cite{Blandford} describe the production of a
coherent electromagnetic pulse by an explosive event in which the entire
mass of the black hole is emitted. If significant numbers of
electron-positron pairs are produced in the event, the relativistically
expanding shell of these particles (a ``fireball'' of Lorentz factor
$\gamma_f$) acts as a perfect conductor, reflecting and boosting the
virtual photons of the interstellar magnetic field. An electromagnetic
pulse results only for $\gamma_f \sim 10^5$ to $10^7$, for typical
interstellar magnetic flux densities and free electron densities. 
Below $\gamma_f \sim 10^5$ the energy emitted by the PBH goes primarily into sweeping up the ambient interstellar plasma, and not into an electromagnetic pulse; above $\gamma_f \sim 10^7$ the number of electron-positron pairs is insufficient to carry the fireball surface current necessary to expel the interstellar magnetic flux density. The energy of the electron-positron pairs is
\begin{equation}
\label{gen1}
kT \approx \frac{\gamma_f}{10^5} \ 0.1 \ {\rm TeV}. 
\end{equation}
Thus the energy associated with $\gamma_f\sim10^5$ corresponds roughly
to the electroweak scale.

\section{Exploding primordial black holes and the TeV scale}

There is a remarkable, heretofore unrecognized, relationship between the
range of pulse-producing Lorentz factors for the emitted particles, and
the TeV scale. Since $\gamma_f \propto T$ at the time of the explosive
burst, equation~(1) yields
\begin{equation}
\frac{\gamma_f}{10^5} \approx \frac{10^{-19}\ \textrm{m}}{R_s},
\label{gen2}
\end{equation}
where $R_s$ is the Schwarzschild radius. Thus, the allowed range of Lorentz factors implies length scales $R_s\sim 10^{-19} - 10^{-21}$~m.
Taking these as 
Compton wavelengths we find the associated energy scales to be
\begin{equation}
(R_s/\hbar c)^{-1}\sim 1-100 \ \textrm{TeV}.
\label{generic2}
\end{equation}
This relationship suggests that the production of an electromagnetic
pulse by PBHs might be used to probe TeV-scale physics. To make use of
this interesting, but fairly generic observation, a specific
phenomenologically relevant explosive process is required. One such
process, which connects quantum gravitational phenomena and the TeV
scale, makes use of the possible existence of an extra dimension.

\section{Explosive primordial black hole evaporation due to the presence of an extra dimension}

Spatial dimensions in addition to the observed 3+1 dimensional spacetime
have a long tradition in gravitational models that goes back to the work
of Kaluza and Klein \cite{Kaluza, Klein}. Extra dimensions are also
required in string/M-theory for the consistency of the theory \cite{GSW}. 
It was traditionally assumed, in these approaches, that the
extra dimensions are Planck length in size. However, various
phenomenologically-motivated models were recently developed with extra
dimensions much larger than the Planck length, which could have
observable implications for electroweak-scale physics
\cite{TeV,UED,ADD,RS1,RS2}.

Black holes in four dimensions are uniquely defined by charge, mass, and
angular momentum. However, with the addition of an extra spatial
dimension, black holes could exist in different phases and undergo phase
transitions. For one toroidally compactified extra dimension, two
possible phases are a black string wrapping the compactified extra
dimension, and a 5-dimensional black hole smaller than the extra
dimension. A topological phase transition from the black string to the
black hole is of first order \cite{Gubser, Cas1, Cas2}, and results in a significant
release of energy equivalent to a substantial increase in the luminosity
of Hawking radiation \cite{Kol2}.

Following the analysis of Kol \cite{Kol1b}, to parametrize the phase of
the black hole we define a dimensionless order parameter $\mu =
GM/Lc^2$, where $L$ is the size of the extra dimension with coordinate $z$
identified with $z+L$. For large values of $\mu$ the black string phase is dominant,
while for small values of $\mu$ the 5-D black hole phase is favored.
PBHs evaporating in the current epoch would lose mass through
evaporation causing $\mu$ to decrease until a metrical instability, the
Gregory-Laflamme point \cite{GL,GL2} ($\mu\approx0.07$) is reached, at
which time the first-order phase transition occurs
\cite{Kol1b,Gubser}.\footnote{While the final state resulting from the 
topological phase transition is not
entirely understood, such details will not significantly alter the
analysis presented here.} The Schwarzschild radius is related to $L$ as
$R_s=2GM/c^2=2\mu L$. Thus, the energy emitted at the topological phase
transition is
\begin{equation}
E = \eta Mc^2 = \eta \frac{R_s c^4}{2G} = \eta \mu L \frac{c^4}{G},
\end{equation}
equivalent to a Planck power (reduced by $\eta\mu$) emitted during a
time scale $L/c$. The factor $\eta$ is an efficiency parameter,
estimated by Kol to be a few percent in analogy with black hole
collision simulations \cite{col}.

\section{Transient pulse production}

The analysis of Rees \cite{Rees} and Blandford \cite{Blandford} can be adapted
to the topological phase transition scenario. For a coherent
electromagnetic pulse to result, the time scale of the energy release
must be $L/c \ll \lambda/c$, where $\lambda$ is the characteristic
wavelength of the pulse. This requirement is well satisfied for the size of 
extra dimension considered here. Since
$\gamma_f \propto T$ and a fraction $\eta$ of the object's mass-energy
is released, the inverse relationship between temperature and mass for
the Hawking process, equation~(\ref{tem}), implies $\gamma_f$ is
inversely related to the energy of the fireball.

Determining the emitted particle spectrum would require a full theory of
quantum gravity. Lacking such a theory, we make the simple assumption
that 50\% of the ejected energy is in the form of electron-positron
pairs (the same assumption used in \cite{Rees,Blandford}).\footnote{The emitted
particle spectrum (and decay chain) for the event considered by Rees and
Blandford, taking into account possible details of the QCD phase
transition, has been investigated \cite{Jane1,Jane2}. However, the
topological phase transition scenario considered here is of a fundamentally
different nature making this analysis inapplicable.} Thus, the energy ejected in electron-positron pairs, and ultimately emitted in the electromagnetic pulse (if $\gamma_f$ is in the appropriate range) is
\begin{equation}
\label{hi}
E_{\rm pulse} \approx E_{{\rm e}^+{\rm e}^-} \approx \eta_{01} \ \gamma_{f5}^{-1} \ 10^{23} \ {\rm J}
\end{equation}
where $\eta_{01} = {\eta}/{0.01}$ and $\gamma_{f5} = \gamma_f/{10^5}$ represent nominal values for these parameters.\footnote{The
nominal value for $\eta$ of 0.01, appearing here and in subsequent equations,
reflects both the few percent efficiency of the topological phase transition, {\it
and} the assumption that 50\% of the released energy is carried away by
electron-positron pairs.}
The bounds on the Lorentz factor for pulse production in the topological
phase transition scenario are of the same order as for the scenario
considered by Rees and Blandford, $\gamma_f \sim 10^5$ to $10^7$.
The size of the extra dimension corresponding to a specific fireball Lorentz factor is
\begin{equation}
L \approx  \mu_{07}^{-1} \ \gamma_{f5}^{-1} \ 10^{-18} \ {\rm m}
\end{equation}
where $\mu_{07} = {\mu}/{0.07}$.

The characteristic frequency of the pulse is 
\begin{equation}
\nu_c \approx \eta_{01}^{-1/3} \ \gamma_{f5}^3 \ b^{2/3} \ 5.1 \ {\rm GHz}
\end{equation}
where $b$ is the interstellar magnetic flux in units of 0.5~nT. Pulses for low $\gamma_f$ are best observed in the radio spectrum. The maximum radius attained by the shell is $\approx R_\odot \eta_{01}^{1/3} \gamma_{f5}^{-1} b^{-2/3}$. The interstellar magnetic field is expected to be essentially uniform on this length scale. Thus a pulse should be nearly 100 percent linearly polarized, which will help to distinguish pulses from PBHs from those produced by other sources.

Following Blandford \cite{Blandford}, the pulse energy spectrum is
\begin{equation}
I_{\nu\Omega} \approx 1.4\times10^{12} \ 
\eta_{01}^{4/3} \ 
\gamma_{f5}^{-4} \ 
b^{-2/3} \ 
\left\vert F\left(\frac{\nu}{\nu_c}\right) \right\vert ^2
\ {\rm J\ Hz}^{-1}\ {\rm sr}^{-1}
\label{spectrum1}
\end{equation}
where the limiting forms of $\vert F(x)\vert^2$ are
\begin{equation}
\left\vert F(x) \right\vert^2 \approx \left\{ \begin{array}{ll}
0.615 \ x^{-4/7} & \textrm{if $x \ll 1$}\\
x^{-4}   & \textrm{if $x \gg 1$}.
\end{array} \right.
\label{spectrum2}
\end{equation}
Equations (\ref{spectrum1}) and (\ref{spectrum2}) imply that for a
chosen observing frequency $\nu$, in GHz, the observed pulse energy
sharply peaks at a specific Lorentz factor,
\begin{equation}
\gamma_{f5} \approx 0.5 \ \eta_{01}^{1/9} \ b^{-2/9} \ \nu_{\rm GHz}^{1/3}.
\end{equation}

By varying the observing frequency, one can search for potential
phase-transition pulses associated with different $\gamma_f$, and thus
different sizes of the extra dimension. The corresponding extra
dimension that is tested for using a particular search frequency has the
size
\begin{equation}
L \approx \mu_{07}^{-1} \ \eta_{01}^{-1/9} \ b^{2/9} \ \nu_{\rm GHz}^{-1/3} \ 2\times10^{-18} \ {\rm m}.
\label{nAndL}
\end {equation}
The strength of the typical interstellar magnetic field varies around the nominal 
value we use by about an order of magnitude \cite{mag}.  For the weak dependence 
of $L$ on $b$ shown in equation (\ref{nAndL}) the resulting error in 
a determination of $L$ is less than a factor of 2.  However, given the idealized nature 
of the Blandford model it is likely that the observations we suggest 
can only determine the size of an extra dimension to an order of magnitude.

Frequencies between $\sim$~1~GHz and $10^{15}$~Hz ($\gamma_f \sim 10^5$
to $10^7$) sample possible extra dimensions between $L\sim
10^{-18}-10^{-20}$~m. These length scales correspond to energies of
$(L/\hbar c)^{-1}\sim0.1-10$~TeV. The electroweak scale is
$\sim$0.1~TeV, and thus, radio observations at $\nu \sim 1$~GHz may be
most significant.

The observed polarization, dispersion measure, and energy of a radio 
pulse would provide a means for distinguishing a PBH explosion from 
other possible sources. As noted above, an electromagnetic pulse 
produced by an exploding PBH would be nearly completely linearly 
polarized, helping to distinguish it from other possible sources. 
In addition, the dispersion measure of a radio pulse can be used 
to estimate the distance to the source of the pulse.  This distance, 
in combination with the observed pulse energy, can yield an emitted 
pulse energy per Hz, at observing frequency $\nu$, that can be 
compared to the expected model results shown in Fig.~1.

The efficiency $\eta$ differs by two orders of magnitude for the 
PBH explosion scenario considered by Rees and the topological phase transition 
scenario.  Therefore, the emitted pulse energy derived from 
observations, for the Lorentz factor probed, 
would distinguish between these two scenarios.  Thus, as Fig.~1 shows, 
given a chosen observing frequency, one can distinguish between the 
cases of $\eta\approx 1$ (all the mass is emitted in a final explosive burst) 
and $\eta\approx 0.01$ (for the topological phase transition).

\begin{figure}
\begin{center}
\includegraphics{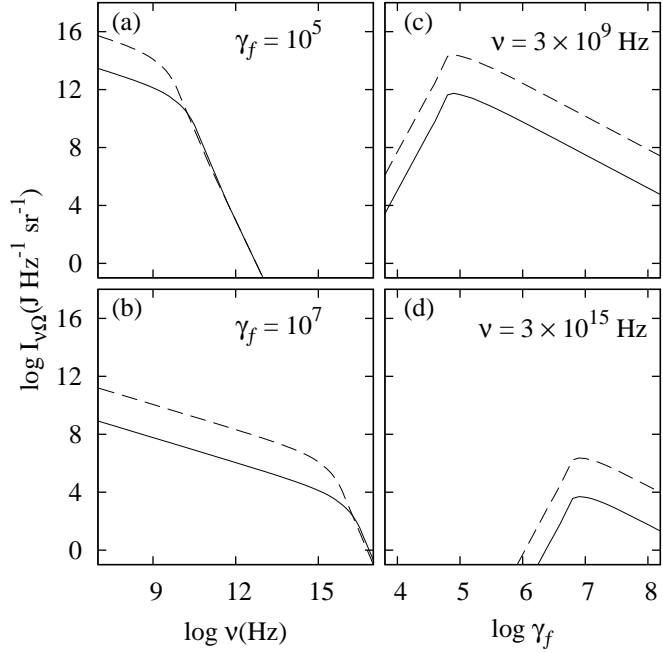}
\caption{
Electromagnetic-pulse energy spectrum for a topological phase
transition (solid curves, efficiency $\eta\approx 0.01$), and a final explosive burst as described by Rees and Blandford (dashed curves, efficiency $\eta\approx1$).
(a) shows the pulse energy per unit frequency interval, versus
frequency, for a fireball Lorentz factor $\gamma_f=10^5$. (b)
is for $\gamma_f=10^7$. (c) shows the pulse energy per unit
frequency interval versus $\gamma_f$, at frequency $3\times10^9$~Hz.
(d) is for frequency $3\times10^{15}$~Hz. Note that the curves
in (c) and (d) peak sharply at a specific $\gamma_f$,
dependent on the observing frequency (full-width at half-maximum
$\approx$ 1/3, in $\log \gamma_f$). So, choosing an observing frequency
enables searching for events of a particular $\gamma_f$. For a
topological phase transition, a $\gamma_f$ is associated with an extra
dimension of a particular size (see text). Pulses for the cases
$\eta\approx0.01$ and $\eta\approx1$ can be distinguished. In (c), for
example, the output energy is dramatically different for the two cases,
at the $\gamma_f$ of peak output. Thus, a distance to the object (e.g.,
from pulse dispersion measure for radio observations) would distinguish
the two cases. A pulse for $\eta\approx1$ can have the same energy as for $\eta\approx0.01$ only for a significantly different $\gamma_f$, which could be determined by sampling the spectrum at multiple wavelengths.
}
\end{center}
\end{figure}

\section{Transient pulse searches}

Searches for transient radio pulses from PBH explosions, cf.
\cite{Phinney,Benz}, can probe for the existence of PBHs well below the
limits established by observations of the diffuse $\gamma$-ray
background \cite{Page,Halzen}. To date, these radio searches have
utilized data collected for other purposes, or for limited times, all
with negative results. A new generation of instruments, designed to
operate at low radio frequencies, may be able to conduct extended
searches for radio transients over wide fields of view ($\sim$ 1
steradian): the Long Wavelength Array (LWA) \cite{LWA}, Murchison
Widefield Array (MWA) \cite{MWA}, and the Low Frequency Array (LOFAR)
\cite{LOFAR}. 

A continuous wide-field low-frequency radio transient search already
underway uses the Eight-meter-wavelength Transient Array (ETA)
\cite{ETA1,ETA2} which operates at 38 MHz using 10 dual-polarization
dipole antennas. ETA observations are most sensitive to $\gamma_f
\approx 10^4$ to $10^5$ ($L \approx 10^{-17}$~m to $10^{-18}$~m). A
second array (ETA2) is under construction at a different site. Comparing
the signals received at both sites will help mitigate radio interference
--- a technique that distinguishes all searches with distributed antenna
arrays from single-antenna searches. This procedure enables the
theoretical sensitivity to be attained. The sensitivity of a radio telescope
to a pulse-producing source is dependent on the temporal broadening of 
an observed pulse due to interstellar scattering and due to dispersion 
across the finite-width frequency channels utilized in the observations. 
Taking account of these effects, the ETA is sensitive 
to transient pulses produced by black-string/black-hole phase transitions out 
to distances of about 300~pc.

It is natural to ask if gamma-ray satellites should have already detected 
an event of the sort we are considering. The Energetic Gamma-Ray Experiment 
Telescope (EGRET) set an upper limit on PBH explosions of 
$<0.05$~pc$^{-3}$~y$^{-1}$  \cite{Fichtel:1994sf}.  This result assumes 
that PBH explosions are of the ``standard'' variety: $\eta\approx1$, 
and occurring with $\gamma_f\sim10^2$, producing a gamma-ray spectrum 
peaking at about 250~MeV, as discussed by Page and Hawking \cite{Page}. 
Given these assumptions, EGRET is sensitive to such events out to distances 
of about 100~pc \cite{Fichtel:1994sf}. If instead one considers outbursts 
due to topological phase transitions with an efficiency of $\eta\approx0.01$, and at $\gamma_f\sim10^2$, the EGRET sensitivity would only be sufficient to observe events 
out to about 10~pc, assuming the same partitioning of output 
energy into gamma-rays and particles. Furthermore, if one is interested in searching for topological phase transition events at the TeV-scale, 
where an extra dimension is more plausible, the outburst energy (proportional to
the mass of the black hole) is an additional factor of $10^3$ smaller, and so the distance is reduced to 0.3~pc.  Moreover, the associated gamma-ray spectrum peaks at 
this much larger energy scale, and outside the energy range of EGRET.  
Therefore, EGRET was not the most suitable instrument for finding the 
topological phase transition events we are considering.

The Fermi Gamma-Ray Space Telescope (formerly GLAST), will observe photons of energies up to about 300~GeV, encompassing energies that would be produced by a topological phase transition at 0.1~TeV.  However, while Fermi is more than an order of magnitude more sensitive than was EGRET \cite{Michelson:2002ng, Baltz:2008wd}, it will be sensitive to these events out to only $\sim$1~pc.

\section{Implications}

Although we have considered a process involving an extra dimension, we
have kept our analysis general in the sense that we have not specified
any particular extra dimension model. We now consider the above proposal
in the context of several specific extra dimension scenarios. 

In the case of TeV-scale compactification models in which all gauge
fields propagate in a single, circular, extra dimension \cite{TeV}, the
current bound on the compactification scale is $(L/\pi\hbar
c)^{-1}\gtrsim6.8$~TeV \cite{PDG}. The Large Hadron Collider (LHC) will
probe these models up to an energy scale of $\sim 16$~TeV. If both gauge
fields and fermions propagate in the extra dimension \cite{UED} the
current bound is $(L/\pi\hbar c)^{-1}\gtrsim300-500$~GeV with the LHC
probing to $\sim 1.5$~TeV \cite{PDG}. Detection of a transient pulse
would imply, as noted above, an extra dimension with $L\sim10^{-18} -
10^{-20}$~m, corresponding to an energy of $\sim 0.1 - 10$~TeV. Thus
constructive comparison of the pulse detection results and LHC results
would be possible.

In the context of the braneworld scenario proposed by Randall and
Sundrum \cite{RS1, RS2} it has been argued that evaporating black holes
will reach a Gregory-Laflamme instability as the radius of the black
hole approaches the AdS radius \cite{champ, emp}. More specifically, in
the Randall-Sundrum I scenario a nominal value of this radius is
10~{TeV}$^{-1}$ \cite{CIN} placing it within the appropriate range for
transient pulse production.

For large extra dimension models \cite{ADD} the effective fundamental
energy scale is much higher than the energy scale of the large extra
dimension $(L/\hbar c)^{-1}$. For a single large extra dimension of size
$L\sim10^{-18} - 10^{-20}$~m the effective fundamental energy scale is
$\sim 10^{10}$~TeV --- much higher than the electroweak scale. Thus,
searches for pulses from topological phase transitions would probe, for
these models, energies inaccessible to accelerator-based approaches for
the foreseeable future.

While a positive pulse detection would signal the existence of an extra
dimension, a null detection would serve to constrain the possible size
of an extra dimension in particular models. Such a constraint
presupposes, of course, the existence of PBHs in abundant enough numbers
to be detectable. These constraints could be strengthened through
consideration of other experimental data, e.g., other types of searches
for PBHs, or cosmological data which further constrain the spectral
index for primordial density irregularities on the appropriate scales,
or accelerator-based searches.

\section{Outlook}

An important avenue for future investigation is the effect more than one
extra dimension would have on transient pulse production. The
nature of this type of topological phase transition for more than one
compact extra dimension is currently under investigation \cite{Kol3}.
Also, the efficiency parameter $\eta$, whose value was estimated above,
can be better determined numerically, which would help to make this
analysis more precise. We have considered a particular explosive event
in the evaporation process of a PBH involving an extra dimension.
However, given the generic relationships noted above,
equations~(\ref{gen1}) and (\ref{gen2}), we believe that a connection
between transient pulse production by PBHs and electroweak-scale physics
is robust beyond the specific analysis present here, and is worthy of
further investigation.

\section*{Acknowledgments}

We gratefully acknowledge George Djorgovski, Bing Feng, Jean Heremans,
Sabine Hossenfelder, Nemanja Kaloper, Barak Kol, Zack Lewis, Jane MacGibbon,
Djordje Minic, Alexey Pronin, Massimo Ricotti, Eric Sharpe, Gregory
Stock, and Roger A. Wendell for insightful discussions and comments.
This work is supported by NSF grant AST--0504677 and by the Pisgah
Astronomical Research Institute.

\bibliographystyle{unsrt}
\bibliography{references}

\end{document}